\documentclass[sigconf]{acmart}
\copyrightyear{2021} 
\acmYear{2021} 
\setcopyright{acmlicensed}\acmConference[WSDM '21]{Proceedings of the Fourteenth ACM International Conference on Web Search and Data Mining}{March 8--12, 2021}{Virtual Event, Israel}
\acmBooktitle{Proceedings of the Fourteenth ACM International Conference on Web Search and Data Mining (WSDM '21), March 8--12, 2021, Virtual Event, Israel}
\acmPrice{15.00}
\acmDOI{10.1145/3437963.3441771}
\acmISBN{978-1-4503-8297-7/21/03}

\usepackage{CJKutf8}
\usepackage{enumitem}
\usepackage{algorithm}
\usepackage{algorithmic}
\usepackage{natbib}
\usepackage{bbm}
\usepackage{amsmath}
\usepackage{amsthm}
\usepackage{subfig}
\usepackage{xcolor}
\usepackage{footnote}
\usepackage{amsmath}

\DeclareMathOperator*{\maximize}{maximize}
\DeclareMathOperator*{\argmax}{argmax}

\newtheorem{innercustomthm}{Theorem}
\newenvironment{customthm}[1]
  {\renewcommand\theinnercustomthm{#1}\innercustomthm}
  {\endinnercustomthm}

\begin{document}
\fancyhead{}

\title{Optimizing Multiple Performance Metrics with Deep GSP Auctions for E-commerce Advertising}

\author{Zhilin Zhang$^{1*}$, Xiangyu Liu$^{1*}$, Zhenzhe Zheng$^2$, Chenrui Zhang$^3$, Miao Xu$^1$, Junwei Pan$^4$,
\and Chuan Yu$^1$, Fan Wu$^2$, Jian Xu$^1$ and Kun Gai$^1$}
\affiliation{ 
  \institution{$^{1}$Alibaba Group, $^{2}$Shanghai Jiao Tong University, $^{3}$Peking University, $^{4}$Yahoo Research
  \and \{zhangzhilin.pt, qilin.lxy, xumiao.xm,yuchuan.yc,xiyu.xj\}@alibaba-inc.com
\and zhengzhenzhe@sjtu.edu.cn, fwu@cs.sjtu.edu.cn
\and chenrui.zhang@pku.edu.cn, pandevirus@gmail.com, jingshi.gk@taobao.com}
}


\renewcommand{\shortauthors}{Zhilin Zhang and Xiangyu Liu, et al.}

\begin{abstract}

In e-commerce advertising, the ad platform usually relies on auction mechanisms to optimize different performance metrics, such as user experience, advertiser utility, and platform revenue. However, most of the state-of-the-art auction mechanisms only focus on optimizing a single performance metric, e.g., either social welfare or revenue, and are not suitable for e-commerce advertising with various, dynamic, difficult to estimate, and even conflicting performance metrics. In this paper, 
we propose a new mechanism called \emph{Deep GSP auction}, which leverages deep learning to design new rank score functions within the celebrated GSP auction framework. These new rank score functions are implemented via deep neural network models under the constraints of \emph{monotone allocation} and \emph{smooth transition}. The requirement of monotone allocation ensures Deep GSP auction nice game theoretical properties, while the requirement of smooth transition guarantees the advertiser utilities would not fluctuate too much when the auction mechanism switches among candidate mechanisms to achieve different optimization objectives.
We deployed the proposed mechanisms in a leading e-commerce ad platform and conducted comprehensive experimental evaluations with both offline simulations and online A/B tests. The results demonstrated the effectiveness of the Deep GSP auction compared to the state-of-the-art auction mechanisms.{\let\thefootnote\relax\footnote{{$^*$Both authors contributed equally to this research.}}}

\end{abstract}



\maketitle

\section{Introduction }
\label{sec:intro}



In e-commerce advertising, the advertisers usually leverage the ad platform to promote their products to their target users and boost the overall merchandise volume~\cite{evans2009online,goldfarb2011online}.
There are three major kinds of stakeholders: users, advertisers, and the ad platform. Users look for good shopping experiences, advertisers want to accomplish their marketing objectives, and the ad platform would like to extract high revenue in providing satisfying services to both users and advertisers.
For the long-term prosperity, one critical tool the ad platform can use to jointly optimize the above mentioned multiple objectives (i.e., performance metrics) is the auction mechanism.
The auction mechanism determines the ads displayed to users as well as the payments charged to advertisers.



The performance metrics of the ad platform can be diverse and the importance of these metrics can vary over time. For example, users would like to find their desired products with small search frictions, requiring the ads displayed to them have high relevance, which is often quantified as Click-Through Rate (CTR) and Conversion Rate (CVR). Advertisers usually want to optimize certain marketing performance metrics such as Gross Merchandise Volume (GMV) under a budget or Return on Investment (ROI) constraint. The ad platform, whose productivity in creating revenue is typically measured by Revenue Per Mille (RPM), also has to provide satisfying shopping experience to users and help advertisers fulfill their marketing objectives. Besides, it is also a common practice for e-commerce ad platforms to change the importance of different metrics over time. For example, the ad platform may put more emphasis on CTR and RPM in ordinary stages, and inclines towards CVR and GMV in shopping festivals to encourage more sales.



The performance metrics from different stakeholders often conflict with each other. One representative example is the conflict between user experience metrics (e.g., CTR) and revenue metrics (e.g., RPM). Simply maximizing RPM may select the ads with high bids but relatively low CTRs and vice-versa. Moreover, optimizing conflicting multiple performance metrics in the ad platform is substantially different from the traditional multi-objective optimization~\cite{lin2019pareto}. As stakeholders have incentives to manipulate the mechanisms for their own interests, the optimization of multiple performance metrics needs to be modeled under a game-theoretic setting. 



Another practical challenge for optimizing multiple performance metrics in e-commerce advertising is that some metrics are difficult to estimate with prediction models. These performance metrics could be related to complicated user interactions. For a displayed ad, a user could click the ad and browse the detailed information. If the user is interested in the product, she would have further actions, \emph{e.g.,} adding it to the shopping cart, and/or placing an order. Besides these complicated user-ad interactions, the performance metrics are also related to many factors in a non-analytical manner. These factors include, but are not limited to the competitions among advertisers and the macro-control strategies of the ad platform. Therefore, some performance metrics, such as RPM and GMV, can only be evaluated by the actual feedback after displaying the ads.

    
While the auction theory provides a rich set of tools for optimizing social welfare or revenue~\cite{lahaie2007revenue,roberts2016ranking,varian2007position}, few of them can be used to optimize the above mentioned diverse, dynamic, conflicting and feedback-based performance metrics. The widely used generalized second-price auction (GSP)~\cite{lahaie2007revenue} selects the ads based on their rank scores, which could be the product of bid and ad quality, where ad quality is usually quantified as predicted click-through rate ($pCTR$) or predicted conversion rate ($pCVR$). However, such a simple and static rank score is not suitable for optimizing multiple performance metrics in e-commerce advertising. Moreover, it is also not clear how to dynamically adjust this rank score for diverse and dynamic optimization objectives.

 

In this paper, we enhance the capability of the celebrated GSP auction with the power of deep learning to optimize multiple performance metrics in e-commerce advertising. We design new rank score functions that map the features such as bid, $pCTR$, $pCVR$, ad category, product price, to a rank score. Different from that in GSP auction, our new rank score functions can be non-linear functions. Such rank score functions can well model the relationship between the rich features and the performance metrics of interest. We implemented these non-linear rank score functions with carefully designed deep neural network models to accommodate various performance metrics. We also require these deep neural network models to be monotone w.r.t. bid, i.e., the auction follows monotone allocation, under which we prove that the Deep GSP auction can achieve nice game theoretical properties. As the importance of different performance metrics can vary over time due to business needs, we further introduce the \emph{smooth transition} constraint to ensure the advertiser performance metrics not fluctuate too much when the auction mechanism switches among candidate mechanisms to achieve different optimization objectives. For the neural network model implementation, we observe that it is not tractable to obtain the auction outcomes in advance as supervision for model training. We therefore make a connection between model training and the exploration process in reinforcement learning, and train the model with a model-free policy optimization algorithm.

The contributions in this paper can be summarized as follows: 

(\romannumeral1) We investigated the new aspects in e-commerce advertising in this paper. We propose an end-to-end learning based ad auction mechanism, namely Deep GSP auction, towards optimizing multiple performance metrics in a dynamic and game theoretical setting. 

(\romannumeral2) We designed deep neural network based rank score functions with the constraints of \emph{monotone allocation} and \emph{smooth transition}. It can be proved that the resulting Deep GSP auction with these rank score functions possesses nice game theoretical properties: \emph{Incentive Compatibility (IC)}~\cite{myerson1981optimal} for single ad slot cases and \emph{Symmetric Nash equilibrium (SNE)}~\cite{varian2007position} for multi-slot cases.

(\romannumeral3) We conducted extensive offline and online experiments to evaluate the effectiveness of Deep GSP auction. The evaluation results demonstrated that the Deep GSP auction outperforms the baseline methods significantly in terms of various performance metrics, including GMV, RPM, CTR, CVR, etc. 

\begin{figure}[!t]
\centering
\includegraphics[scale=0.45]{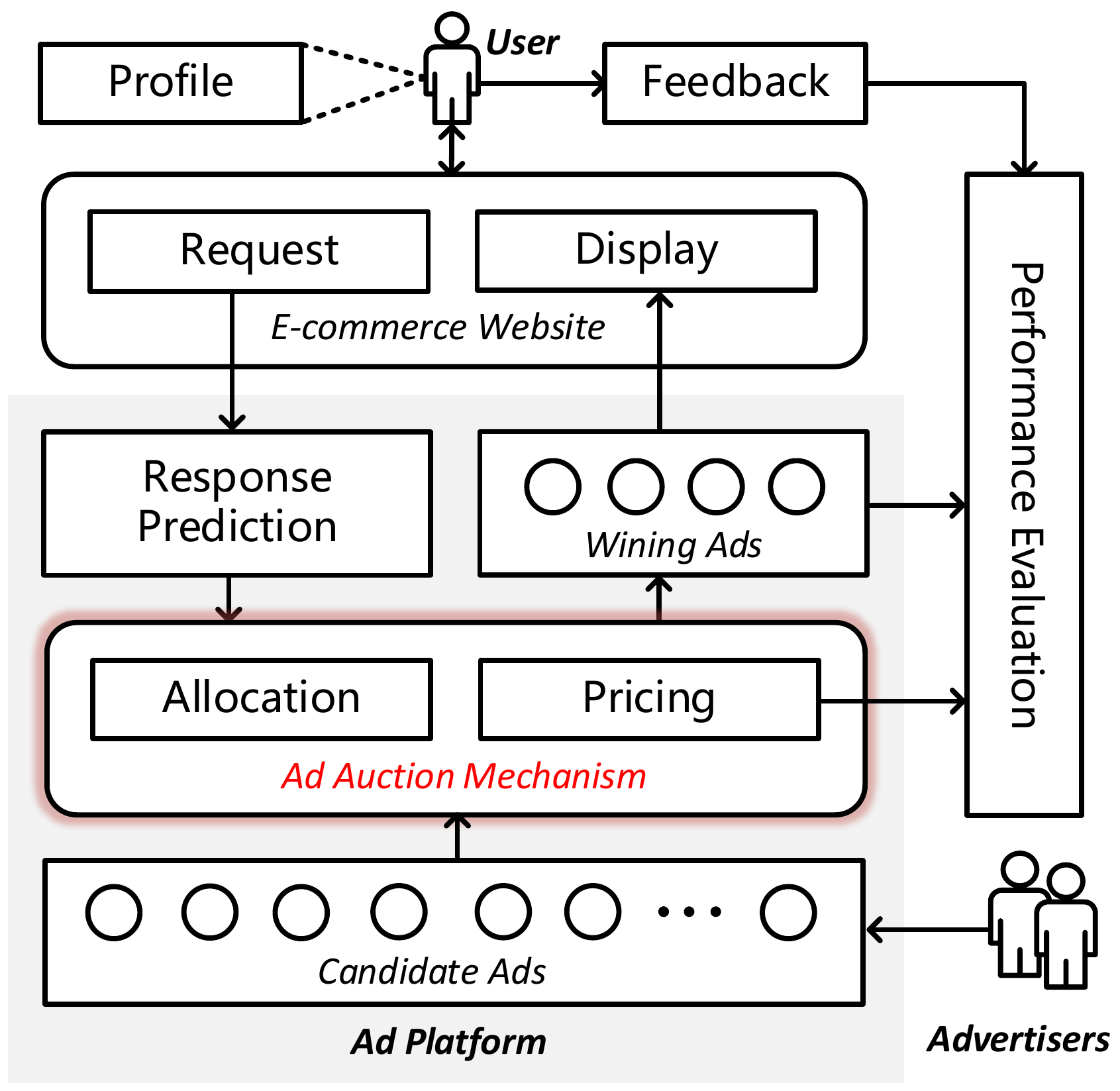}
\caption{An ad platform architecture in e-commerce advertising. First, a user visits the e-commerce website, which sends a request to the ad platform. Secondly, advertisers perform bidding upon the candidate ads according to the attributes of the request. Inside the ad platform, the mechanism module selects top-N ads and charges corresponding payment through the allocation and pricing module, respectively. These winning ads will be displayed to users. Finally, the user may have a series of interactions with these ads, which are the performance metrics concerned in this paper.}
\label{fig:system}
\end{figure}

\section{Preliminaries}
\label{sec:preliminaries}
\subsection{Ad Platform Architecture}
\label{sec:system_model}

As shown in Fig.~\ref{fig:system}, we describe a typical ad platform architecture in e-commerce advertising. There are $N$ advertisers compete for $K \leq N$ ad slots, which are incurred by an ad request from the user. Each advertiser $i$ submits bid $b_i$ to participate in the auction, usually based on her private valuation $v_i$. We note that $b_i$ does not necessarily equal to $v_i$. We use vector $\mathbf{b}=(b_i,\mathbf{b}_{-i})$ to represent the bids of all the advertisers, where $\mathbf{b}_{-i}$ represents the bids from all the advertisers except $i$. The response prediction module predicts the user response probabilities  (e.g., $pCTR$, $pCVR$, etc.) on the candidate ads. These predictions, together with the bids $b$, could be used by the auction mechanism module, denoted by $\mathcal{M}\langle\mathcal{R},\mathcal{P}\rangle$, for top-$K$ ads selection and payment calculation.
We use $\mathcal{R}_i(b_i, \mathbf{b}_{-i})=k$ to denote that the advertiser $i$ wins the $k$-$th$ ad slot, while $\mathcal{R}_i(b_i, \mathbf{b}_{-i})=0$ represents the advertiser loses the auction. The $K$ winning ads would be displayed to the user. The auction mechanism module further calculates the payment for the winning ads following a payment rule $\mathcal{P}$.
Let $\mathcal{P}_i$ be the payment charged to advertiser $i$, and thus her utility is $u_i  = v_i -\mathcal{P}_i$. Finally, the performance metrics can be obtained from the payments of the displayed ads and the responses from users after seeing the ads.
In this paper, we focus on designing an ad auction mechanism to optimize $L$ performance metrics, which can be expressed as functions ${\{f_j(\mathbf{b};\mathcal{M})\}}_{1}^{L}$. We explicitly show the performance metrics depend on the bidding profile and the deployed auction mechanism.  

\subsection{Problem Formulation}
\label{sec:problem_formulation}
Based on the concepts discussed above, we formulate the considered problem as \emph{multiple performance metrics optimization in the competitive advertising environments}. Given bid vector $\textbf{b}$ from all the advertisers and $L$
ad performance metric functions ${\{f_j(\mathbf{b};\mathcal{M})\}}_{1}^{L}$, we aim to design an ad auction mechanism $\mathcal{M}\langle \mathcal{R}, \mathcal{P} \rangle$, such that

\begin{equation}
\begin{aligned}
\maximize_{\mathcal{M}} \quad & \mathbb{E}_{\mathbf{b} \sim \mathcal{D}} \left[\sum_{j=1}^L w_j \times f_j(\mathbf{b};\mathcal{M})\right] \\
\textrm{s.t.} \quad 
& \textit{Game Equilibrium constraints,}\\
& \textit{Smooth Transition constraints,}\\
\end{aligned}
\label{eq:problem}
\end{equation}
where $\mathcal{D}$ is the advertisers' bid distribution based on which bidding vectors $\mathbf{b}$ are drawn. 
The objective is to maximize a linear combination of the multiple performance metrics $\{f_j\}^L_1$ with preference parameters. By choosing different $w_j$'s, we can design auction mechanisms to make various trade-offs among performance metrics. In this paper, we assume $w_j$'s are the inputs in the problem formulation, and focus on the ad auction mechanism design. There are extensive related works on how to determine $w_j$'s and derive Pareto-efficient solutions~\cite{lin2019pareto,chen2017optimizing}.
We consider some desirable properties, i.e. \textit{Game Equilibrium} and \textit{Smooth Transition}, when designing an ad auction mechanism. 
There are several equilibrium concepts from game theory for auction design, e.g., Nash Equilibrium (NE)~\cite{varian2007position} and Incentive Compatibility (IC)~\cite{myerson1981optimal}.
IC is a desired economic property for auction design in competitive environments. Intuitively, an ad auction mechanism is IC, if all the bidders truthfully reveal their private valuations. The IC mechanism would remove the burden of considering bidders' strategic behaviors, leading to reliable and predictable inputs for ad performance optimization. Therefore, the IC auction mechanisms could promote the long-term prosperity of the advertising ecosystem~\cite{hammond1979straightforward}.




The most well-known sufficient condition for an IC auction mechanism is the classical Myerson Theorem~\cite{myerson1981optimal}. We first present this condition in the context of single slot ad auction. 
\begin{customthm}{1}[\protect{\cite{myerson1981optimal}}]\label{thm:IC}
A single slot auction mechanism $\mathcal{M}\langle \mathcal{R}, \mathcal{P} \rangle$ is  \textit{incentive-compatible} if and only if the allocation scheme $\mathcal{R}$ is monotone, i.e., the winning bidder would still win the auction if she reports a higher bid, 
and the pricing rule is based on the critical bid, which is the minimum bid that the winning bidder needs to report to maintain the winning state:
\begin{equation}\label{eq:IC_mono}
    \mathcal{R}_i(z, \textbf{b}_{-i}) \geq \mathcal{R}_i(b_i, \textbf{b}_{-i})\mbox{ if  } z > b_i\mbox{  (Monotone Allocation)}
\end{equation}
\begin{equation}\label{eq:IC_mp}
    \mathcal{P}_i = inf_{z|\mathcal{R}_i(z, \textbf{b}_{-i})=\mathcal{R}_i(b_i, \textbf{b}_{-i})}\mbox{  (Critical Bid based Pricing )}
\end{equation}
\qed
\end{customthm}

For the multi-slot case, we turn to a widely used solution concept in the ad industry: Symmetric Nash Equilibrium (SNE).

\footnotetext[1]{Here the notations of  $i$ and $j$ are slightly abused.}

\begin{customthm}{2}[\protect{\cite{varian2007position}}]\label{thm:SNE}
An auction mechanism $\mathcal{M}\langle \mathcal{R}, \mathcal{P} \rangle$ satisfies symmetric Nash equilibrium (SNE) if and only if each bidder in this equilibrium prefer her current allocated slot $i$ to any other slot $j$:
\begin{equation}\label{eq:SNE}
 \beta_i(v_i-p_i) \ge \beta_j(v_i-p_j),
\end{equation}
where $\beta_i$ is the inherent click-through rate of the slot $i$\footnotemark[1].\qed
\end{customthm}

Another desirable property we want to achieve is \emph{Smooth Transition (ST)}.
As discussed in Section \ref{sec:intro}, the performance objectives of the ad platform may vary due to the change of the business logic. If the new optimization objective is quite different from the previous one, the resulting auction mechanism would significantly affect the advertisers' utilities ~\cite{bachrach2014optimising}. 
This introduces the chaos of the auction environment. 
To stabilize advertisers' utility change under different mechanisms, we choose a benchmark mechanism $\mathcal{M}_0$, and require advertisers' utility under the new mechanism should not be less than $1-\epsilon$ of that under  $\mathcal{M}_0$.
The benchmark mechanism $\mathcal{M}_0$ could be the currently deployed mechanism.
Specifically, we define the \textit{Smooth Transition constraint} as follows:
\begin{equation}
u_i (\mathcal{M}) \geq (1 - \epsilon) \times \bar{u}_i (\mathcal{M}_0),    
\end{equation}
where we set a lower bound for advertiser $i$'s utility ${u}_i$ when selecting a new auction mechanism $\mathcal{M}$. 
The lower bound $\bar{u}(\mathcal{M}_0)$ could be set as the average utility over a certain period under the benchmark mechanism $\mathcal{M}_0$.
The parameter $\epsilon$ is a tolerant utility loss ratio for advertisers ($0 \le \epsilon \le 1$). 
By choosing an appropriate $\epsilon$, the advertiser's utility would not fluctuate too much when the auction mechanism is switched towards optimizing another objective.



Classical approaches from mechanism design usually resort to the reliable prediction model for performance metrics, and search for the solutions by simply maximizing the expected performance objective~\cite{bachrach2014optimising}. 
However, precise predictions are intractable, especially when some performance metrics are related to the long sequential user interactive behaviors. 

\section{Deep GSP Auction}
\label{sec:methodology}
In this section, we introduce a \emph{Deep GSP auction} with a deep neural network model to map all the related features to a rank score. The deep neural network model is optimized towards the preferred performance objective. Then we interpret the optimization problem in Eq. (\ref{eq:problem}) as a decision-making problem, and solve it with deep policy optimization. Our approach benefits from the expressive power of deep neural networks and the ability to enforce the aforementioned constraints in training using the standard decision-making pipeline.

\subsection{Deep GSP Auction Design}
\label{sec:model_IC}
We follow the design rationale of the classical GSP auction mechanism~\cite{lahaie2007revenue}, where the allocation scheme is to rank advertisers according to their \emph{rank scores} with a non-increasing order, and the payment rule is to charge the winning advertiser with the minimum bid required to maintain the allocated rank position. The rank score of the  classical GSP auction is the product of the bid and the ad \emph{quality}, which can only optimize certain performance metrics, such as social welfare or revenue. 
To optimize multiple ad performance metrics, we leverage the deep learning technique to design a new rank score and integrate it into the GSP auction framework. We call this new mechanism \emph{Deep GSP auction}. Specifically, we design a deep neural network to map advertiser's bid to a rank score, with the consideration of various related information, such as ad features (ads category, $pCTR$, and $pCVR$), user profile (gender, age, and income) and advertiser preference (budget, marketing demands). 
We use $r_i = R_{\theta}(b_i, \textbf{x}_{i})$ to denote this new rank score, where $\textbf{x}_{i}$ represents the related features except the bid. 
The training of this deep rank score model is under the guideline of the optimization objective in problem (1). In order to satisfy the game equilibrium constraint, we also require the mapping function $R_{\theta}(b_i, \textbf{x}_{i})$ to be monotone with respect to the bid $b_i$. We would discuss how to train the model to satisfy this property later.  

With this new rank score, the allocation scheme and payment rule in Deep GSP auction can be summarized as follows:


\begin{itemize}
    \item Allocation Scheme $\mathcal{R}$: Advertisers are sorted in a non-increasing order of new rank score $r_i = R_\theta(b_{i},\textbf{x}_{i})$: 
    $$r_1\geq r_2 \geq \cdots \geq r_N.$$ 
    The advertisers with the top-K scores would win this auction. Ties are broken arbitrarily.  
    \item Pricing Rule $\mathcal{P}$. The payment for the winning advertiser $i$ is calculated by the formula:
    \begin{equation}\label{eq:pricing}
    p_{i}=R_{\theta}^{-1}(r_{i+1}, \textbf{x}_{i}),
    \end{equation}
    where $r_{i+1}$ is the rank score of the next highest advertiser, and $R_{\theta}^{-1}(\cdot, \textbf{x}_{i})$ is the inversion function of $R_{\theta}(\cdot, \textbf{x}_{i})$.
\end{itemize}




The remaining question is how to train the deep neural network model with the monotone property, and how to efficiently calculate the inverse operation in payment rule.

\subsubsection{Point-wise Monotonicity Loss}
\label{sec:PML}
In principle, monotonicity with respect to a certain subset of inputs can be guaranteed by designing specific neural network architectures. Several pieces of previous work have explored this direction, either from positive weight constraints~\cite{dugas2009incorporating},
or new network architectures~\cite{you2017deep}. However, these methods increase the computational complexity of the training procedure, and may have poor scalability when deployed in a large ad platform. 
In contrast, we directly incorporate the monotonicity constraint within the model training process, and introduce a point-wise monotonicity penalty term into the loss function:
\begin{equation}
\label{eq:mono_loss}
\mathcal{L}_{mono} =\sum_{i=1}^{N} \max(0, -\nabla_{b}R_{\theta}(b_{i},\textbf{x}_{i})), 
\end{equation}
where $\nabla_{b}R_{\theta}(b_{i},\textbf{x}_{i})$ is the gradient of $R_{\theta}$ with respect to bid $b_{i}$. The monotone property implies $\nabla_{b}R_{\theta}(b_{i},\textbf{x}_{i}))\geq 0$. 
This approach is independent of the model structure, which facilitates seamless integration with the already deployed models, and therefore preserves the versatility of deep network architectures. 

\subsubsection{Approximate Inverse Operation}
\label{sec:approximate_inv}
From Eq. (\ref{eq:pricing}), we can precisely derive the payment for each winning advertiser by inverting the rank score function. However, this approach needs to compute complicated pseudo-inverse matrices layer-by-layer in the deep neural network, which can be messy when the weight matrices are ill-conditioned, \emph{e.g.,} singular. As the rank score function $\mathcal{R}_{\theta}$ is monotone in terms of the bid, one potential solution is to use binary search to find the \emph{critical bid}. However, for each binary comparison, we need to re-run the deep neural network with a new bid, which would incur high computing time and is not tractable in online ad auctions (usually needs to complete the payment process within ten milliseconds.). 
To reduce computational complexity and inspired from the payment rule in GSP auction, we propose an approximate inverse operation. We first decompose the rank score $R_{\theta}(b_i,\textbf{x}_{i})$ with a bid multiplier:
\begin{equation}\label{eq:rank_score_bidcoef}
    r_{i}=R_{\theta}(b_i,\textbf{x}_{i})  = b_i \times  \pi_\theta(b_{i}, \textbf{x}_{i}), 
\end{equation}
where $\pi_\theta(b_{i}, \textbf{x}_{i})$ is a non-linear function with bid, and is modeled by a deep neural network. Under this decomposition, we renew the monotonicity penalty term from Eq. ~(\ref{eq:mono_loss}):
\begin{equation}\label{eq:mono_loss_bidcoef}
\mathcal{L}_{mono} = \sum^{N}_{i=1} \max(0, -(\pi_{\theta}(b_i, \textbf{x}_{i}) + b_i\nabla_{b}\pi_{\theta}(b_i,\textbf{x}_{i}))).
\end{equation}
We have observed from an industrial data set that the non-linear function $\pi_{\theta}(b_i, \textbf{x}_{i})$ is not so sensitive to the bid (please refer to the experiment results in Section~\ref{sec:exp_offline_mono_per_check} for more details). 
Thus, in payment calculation, we regard $\pi_\theta(b_i, \textbf{x}_i)$ as a constant w.r.t. $b_i$, similar to the ad quality score in GSP auction, and approximate the payment of the winning advertiser $i$ as:
\begin{equation}\label{eq:pricing_approximated}
    p_i = \frac{r_{i+1}}{\pi_\theta(b_i, \textbf{x}_{i})}.
\end{equation}
With this payment calculation scheme, we do not need to compute the matrix inversion and re-run the neural network. However, the potential drawback of this approximation is that the IC property could not be strictly satisfied. Nevertheless, in Section \ref{sec:exp_offline_mono_per_check}, empirical studies based on an industrial data set demonstrates the effectiveness of this approximation in reducing the computational complexity, and shows that the advertisers could only obtain a limited additional utility.




\subsubsection{Discussion}
In this part, we provide an in-depth analysis of the Deep GSP auction. We demonstrate that the proposed mechanism can be extended to multi-slot auction and has a positive effect to advertising ecological health.

Previous discussions about the IC property are mainly centered around single slot auction. Now we formulate the game equilibrium property of the Deep GSP auction in multi-slot case.

\begin{customthm}{3}\label{thm:DGSP_SNE}
There exists a non-empty set of Symmetric Nash Equilibrium (SNE) states in the Deep GSP auction.\qed
\end{customthm}
Due to space limit, we put the proof of Theorem \ref{thm:DGSP_SNE} in the supplementary material~\cite{Anonymous2020supplementary}.

In addition to optimizing the given objective, Deep GSP also has an incentive effect on advertisers to optimize their ads' quality. Here we give an example to help understand this.

\begin{example}
Suppose that there are three eligible ads, and two ads need to be selected to display. The advertising mechanism wants to optimize both revenue and CTR. Table \ref{tab:gsp_example1} gives the auction result within GSP. According to the $eCPM$ ranking in the GSP, i.e. $pCTR \times bid$, \textit{Ad 1} and \textit{Ad 2} are picked out to display. We manually set a non-linear rank score function, $r = (bid / 10.0)^{0.4} \times (pCTR / 1.0)^{0.7}$, to demonstrate the potential of Deep GSP, whose auction result is shown in Table~\ref{tab:gsp_example2}. Note that the payment for winning ads (PPC) is calculated by second-price for the GSP and Eq.~(\ref{eq:pricing_approximated}) for Deep GSP, respectively.

\begin{table}[!t]
  \centering
  \caption{An example: Three eligible ads and their pCTRs, bids, and two of them will be selected under different mechanisms. Deep GSP outperforms GSP on both total revenue and CTR from the following auction result.\protect\footnotemark[2]}
\small
  \subfloat[GSP auction\label{tab:gsp_example1}]{
    \begin{tabular}{ccc|ccccc}
    \hline
    Ad \# & Bid & pCTR & eCPM & Rank & PPC & Revenue & CTR \\
    \hline
    1 & 10 & 0.1 & 1 & 1 & 4.8 & 0.48 & 0.1 \\
    2 & 2.4 & 0.2 & 0.48 & 2 & 1.95 & 0.39 & 0.2 \\
    3 & 1.3 & 0.3 & 0.39 & 3 & / & / & / \\
    \hline
    \multicolumn{3}{c}{Total} & & & & 0.87 & 0.3 \\
    \hline
    \end{tabular}%
  }
  \\
  \subfloat[Deep GSP (with pre-defined non-linear rank score function)\label{tab:gsp_example2}]{
    \begin{tabular}{ccc|ccccc}
    \hline
    Ad \# & Bid & pCTR & Score & Rank & PPC & Revenue & CTR \\
    \hline
    1 & 10 & 0.1 & 0.199 & 1 & 9.54 & 0.954 & 0.1 \\
    2 & 2.4 & 0.2 & 0.183 & 3 & / & / & / \\
    3 & 1.3 & 0.3 & 0.190 & 2 & 1.25 & 0.375 & 0.3 \\
    \hline
    \multicolumn{3}{c}{Total} & & & & \textbf{1.329} & \textbf{0.4} \\
    \hline
    \end{tabular}
  }
\label{tab:gsp_example}
\vspace{-2.1em}
\end{table}




\begin{figure*}[!t]
\centering
\includegraphics[scale=0.28]{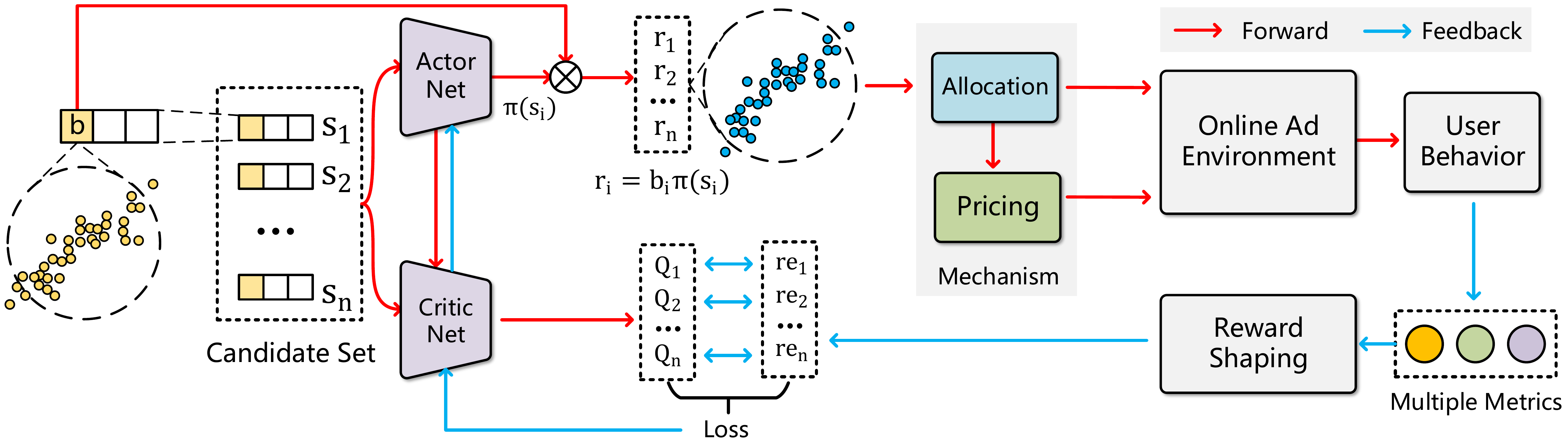}
\vspace{-0.5em}
\caption{A reinforcement learning based framework to implement \emph{Deep GSP auction}. The actor net takes states $\{s_i\}_{i=1}^{n}$ from the candidate set as input for calculating rank scores $\{r_i\}_{i=1}^{n}$. The monotone correlation between the bid dimension of states  $\{s_i\}_{i=1}^{n}$ and rank scores $\{r_i\}_{i=1}^{n}$ is guaranteed by the rank score function $\mathcal{R}_{\theta}$.
Based on the rank scores $\{r_i\}_{i=1}^{n}$, the allocation and pricing modules are deployed in the ad environment, and cooperate with the critic net for multiple metrics optimization.}
\vspace{-0.5em}
\label{fig:framework}
\end{figure*}

We observe that Deep GSP outperforms GSP on both total revenue and CTR. The non-linear score function encodes a sophisticated rank rule, which strikes a favorable balance between two metrics, leading to overall improvements on both metrics. 
Compared with GSP, Deep GSP encourages \textit{Ad 3} to be displayed, who has the highest CTR, despite its eCPM is lower than \textit{Ad 2}. Although \textit{Ad 1} has the highest expected revenue (eCPM), it must pay more to maintain its display, as it has the lowest CTR and may cause poor user experience. This encourages advertisers to optimize their ads' quality to promote CTR and furthermore to improve the user experience, which has a positive effect on the ecological health of the entire auction ecosystem. It is also worth noting that although the payment in Deep GSP is close to the First Price (FP) auction in this example, there is a significant difference between Deep GSP and FP auction. Our proposed mechanism can be proved to be incentive compatible in single slot auction and satisfies SNE in multi-slot auction.\qed
\end{example}



\footnotetext[2]{In this toy example, we assume $CTR$ equals to $pCTR$, and $Revenue$ can be derived from $PPC \times pCTR$.}



\subsection{Deep GSP Auction Implementation}


As introduced in Section \ref{sec:problem_formulation}, some performance metrics are not feasible to have rigorous mathematical analyses, and we can only evaluate these metrics via actual feedback from the system after deploying the auction mechanism. This phenomenon is similar to the exploration process in reinforcement learning, where we need to conduct actions to observe the actual reward. With this connection, we formulate the training of the deep rank score model as a reinforcement learning problem and solve it via model-free policy optimization. 
As shown in  Fig.~\ref{fig:framework}, we introduce a concrete optimization framework to implement the Deep GSP auction and illustrate the detailed procedure.
Given bid vector $\textbf{b}$ of all the advertisers drawn from $\mathcal{D}$, we define the concepts of state, action, reward, and transition. 

\begin{itemize}
    \item \textbf{State:} The state $s_i$ would reflect the quality of ad opportunity and the status of auction environments. We consider the following information to represent state: 1) Ad information, such as bid, $pCTR$, $pCVR$, and ad category. 2) Advertisers' information, like the current budget, the price of products, and marketing intent. 3) User features, such as gender, age, income level, shopping preferences, and etc. Concretely, for ad $i$, $s_i$ equals to $(b_i, \mathbf{x_i})$, which is defined in Section~\ref{sec:model_IC}.
    \item \textbf{Action:} The action $a_i$ is the outcome of the deep rank score model with the state $s_i$ as input, that is the rank score $r_i$ in Eq.~(\ref{eq:rank_score_bidcoef}).
    \item \textbf{Reward:} After taking actions (obtaining rank scores), we run Deep GSP auction and observe performance metrics' realization. We calculate the linear combination of these performance metrics and apply reward shaping strategy to incorporate the \textit{ST constraint} into the reward, which is incurring a large penalty coefficient $ \eta $ if this constraint is violated: 
\begin{equation}
\label{eq:reward_func}
    \quad\quad\quad re_i = F(\mathbf{b};\mathcal{M}) -\eta \times  \max(0, (1 - \epsilon) \times \bar{u}_i(\mathcal{M}^0) - u_i(\mathcal{M})),
\end{equation}
where $F(\mathbf{b};\mathcal{M}) = \sum_{j}w_j\times f_j$ is the optimization objective, i.e. performance metrics defined in Eq.~(\ref{eq:problem}).
We assume all the ads, no matter winning the auction or not, contribute to the auction outcome equally and share this global reward.
    \item \textbf{Transition:} Since our model training is a single-step decision making problem, we can derive the policy without considering the transition dynamics.
\end{itemize}

The goal is to learn an optimal rank score policy $R_{\theta}^*$, which is defined in Eq.~(\ref{eq:rank_score_bidcoef}), that maximizes the expected reward, i.e., 
\begin{equation}\label{eq:DDPG_target}
    R_{\theta}^*=\argmax_{R_{\theta}}\mathbb{E}_{\mathbf{b} \sim \mathcal{D}}[re_i|R_{\theta}].
\end{equation}

Since the optimization algorithm is orthogonal to our proposed Deep GSP auction, any continuous policy optimization method can be used. In this paper, we adopt an actor-critic based approach: Deep Deterministic Policy Gradient (DDPG)~\cite{lillicrap2015continuous}. 
In our context, the update rules of critic and actor model in DDPG are as follows:
\begin{align}
y_{i} &= re_{i}, \label{eq:DDPG_Qtarget}\\
\mathcal{L}(Q_{\theta}) &= \frac{1}{N}\sum_{i}(y_{i}-Q_{\theta}(s_{i},a_{i}))^2, \label{eq:DDPG_QLoss}\\
\mathcal{L}(R_{\theta}) &= \frac{1}{N} \sum_{i} (-Q_{\theta}(s_{i}, R_{\theta}(s_{i})) + \gamma \times \mathcal{L}_{mono}), \label{eq:DDPG_ALoss}
\end{align}
where $\gamma$ is a hyperparameter that modulates the monotonicity constraint applied to the deep rank score model (i.e. actor model).

Our algorithm is different from the vanilla DDPG in two aspects. i) As we formulate the optimization of Deep GSP as a single-step decision making problem, the ground truth for critic $Q$ in Eq.~(\ref{eq:DDPG_Qtarget}) is the reward function solely, in contrast to the bootstrapping form of next state value estimation using target network~\cite{lillicrap2015continuous}. This allows us to pre-train a critic model with log data simply by regression, which extremely improves the sample efficiency. ii) Besides the policy gradient, the actor model training also contains a point-wise monotonicity loss in Eq.~(\ref{eq:DDPG_ALoss}). However, these variations do not prevent us from using the standard training pipeline of DDPG~\cite{lillicrap2015continuous}.

\section{Experimental Evaluations}
\label{sec:exp}


In this section, we firstly introduce the experiment setup, including the evaluation metrics and baselines. Then we conduct offline simulations to evaluate 1) the performance comparison with baseline mechanisms, 2) the effectiveness of point-wise monotonicity loss and approximate inverse payment, 3) the IC/ST properties. Finally, we deploy the deep GSP mechanism on a real e-commerce ad platform, and collect the evaluation results.

\subsection{Experiment Setup}
\subsubsection{Evaluation Metrics}
\label{sec:evaluation_metrics}
We consider the following metrics in our offline and online experiments, which reflect the platform revenue, advertisers' utility, as well as user experience in the e-commerce advertising. For all the experiments in this paper, metrics are scaled to $[0,1]$, without loss of generality.


\textbf{1) Revenue Per Mille (RPM).}
$RPM = \frac{\sum click\times PPC}{\sum impression}\times1000$.


\textbf{2) Click-Through Rate (CTR).}
$CTR = \frac{\sum click}{\sum impression}$. 


\textbf{3) Add-to-Cart Rate (ACR).}
$ACR = \frac{\sum add\text{-}to\text{-}cart}{\sum impression}$.


\textbf{4) Conversion Rate (CVR).}
$CVR = \frac{\sum order}{\sum impression}$.

\textbf{5) GMV Per Mille (GPM).}
$GPM = \frac{\sum merchandise\mbox{ }volume}{\sum impression} \times 1000$.


Apart from the advertising indicators, we also evaluate the economic properties of the proposed Deep GSP auction.

\textbf{6) Monotonicity Metric ($\mathbf{\mathcal{T}_{m}}$).} To verify the point-wise monotonicity proposed in Section \ref{sec:PML}, we conduct experiments on all test data by uniformly generating a set of bids (while keeping other features unchanged) and feeding all the augmented test data set into the trained deep rank score model. We calculate the \textit{Spearman's rank correlation coefficient} ($\rho$)~\cite{corder2014nonparametric} between all the generated bids and their corresponding model outputs to measure the monotonicity of deep rank score model:
\begin{equation}
\mathcal{T}_{m} = \frac{1}{n}\sum\rho_{rank_{bids}, rank_{outputs}}, 
\end{equation}
where $n$ is the size of test data. The $\mathcal{T}_{m}$ takes a range of values from $-1$ to $+1$, where $+1$ suggests a strong monotonic increasing relation.


\textbf{7) Payment Error Rate (PER).} As described in Section \ref{sec:approximate_inv}, the approximate payment solution would introduce an error. In offline simulation, we can use binary search to find the precise payment ($p_i^*$). We denote the $\frac{p_i}{p^*_i}$ by payment error rate (PER). 

\textbf{8) Incentive Compatibility (IC).}
We leverage a data-driven metric, Individual Stage-IC (i-SIC)~\cite{deng2020data}, to quantify IC property. Let $\hat{u}(b) = b \times x(b) - p(b)$ be the advertiser's utility assuming she reports truthfully, where $x(\cdot)$ is the allocation probability. Then the i-SIC metric for a advertiser with valuation $F$ in a mechanism is defined as:
\begin{equation}
    \mbox{i-SIC} = \lim_{\alpha \to 0}\frac{\mathbb{E}_{v \sim F}[\hat{u}((1+\alpha)v)] - \mathbb{E}_{v \sim F}[\hat{u}((1 - \alpha)v)]}{2\alpha \times \mathbb{E}_{v \sim F}[v\times x(v)]}.
\end{equation}
i-SIC metric simply applies small perturbations to bids and records the resulting bidder utilities to quantify IC, which can be computed by straightforward black-box simulations over auction logs.

\subsubsection{Baseline Methods}
We compare Deep GSP with the widely used mechanisms in the industrial advertising environment.

\textbf{1) Generalized Second Price auction (GSP).} In the GSP framework, ads are sorted by expected Cost Per Milles (eCPM). The payment rule for a bidder is the value of the minimum bid required to retain the same slot. The work~\cite{lahaie2007revenue} suggested incorporating a squashing exponent $\sigma$ into the rank score function, i.e. $bid \times pCTR^{\sigma}$ could improve the advertising performance, where $\sigma$ can be adjusted to weight the performance of revenue and CTR. We will refer to this exponential form as GSP.


\begin{figure}[!htbp]
\vspace{-0.6em}
  \centering
  \subfloat[CTR \& RPM]{\label{fig:offline_CTR_RPM}\includegraphics[scale=0.25]{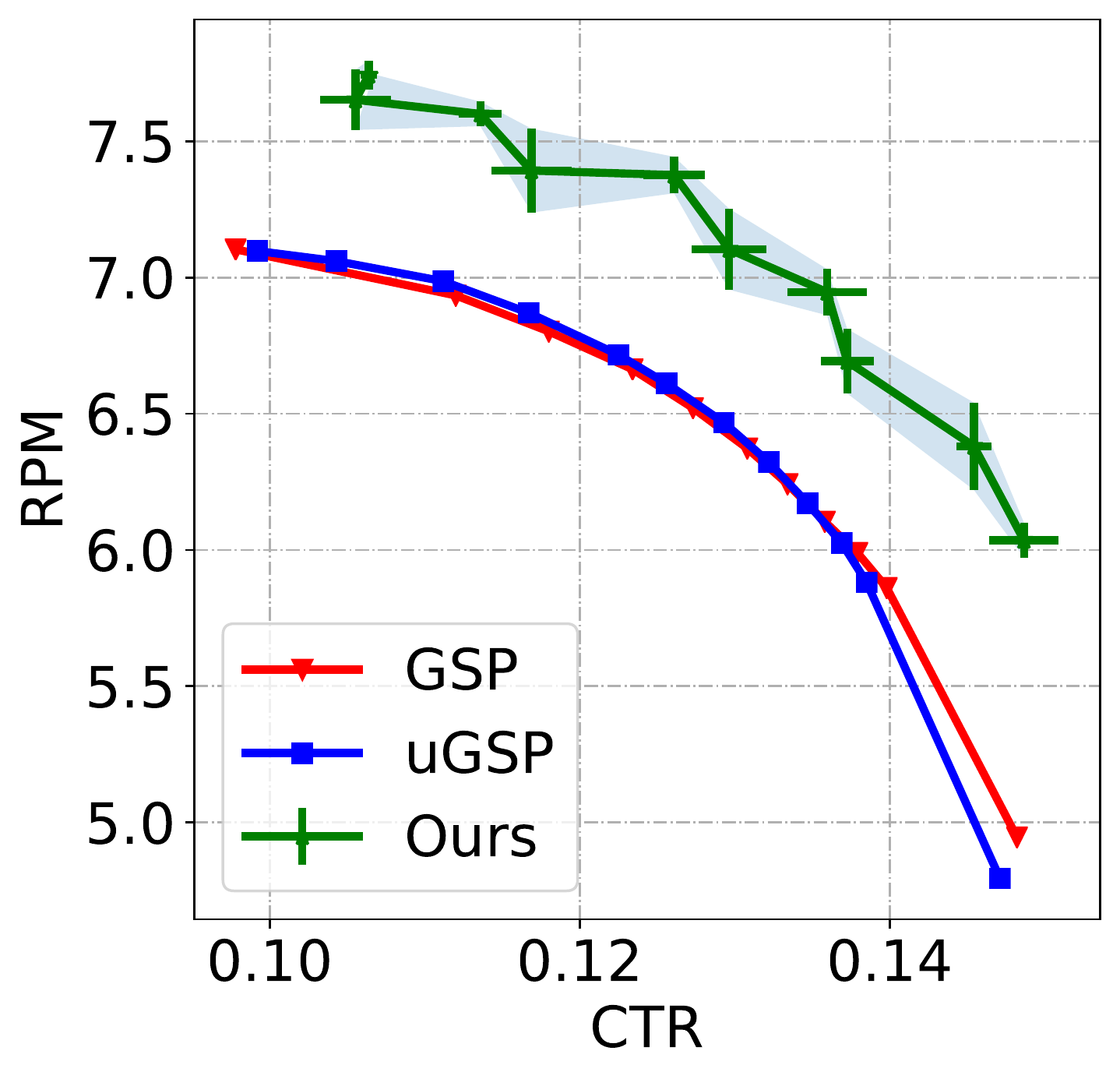}}
  \subfloat[ACR \& RPM]{\label{fig:offline_CR_RPM}\includegraphics[scale=0.25]{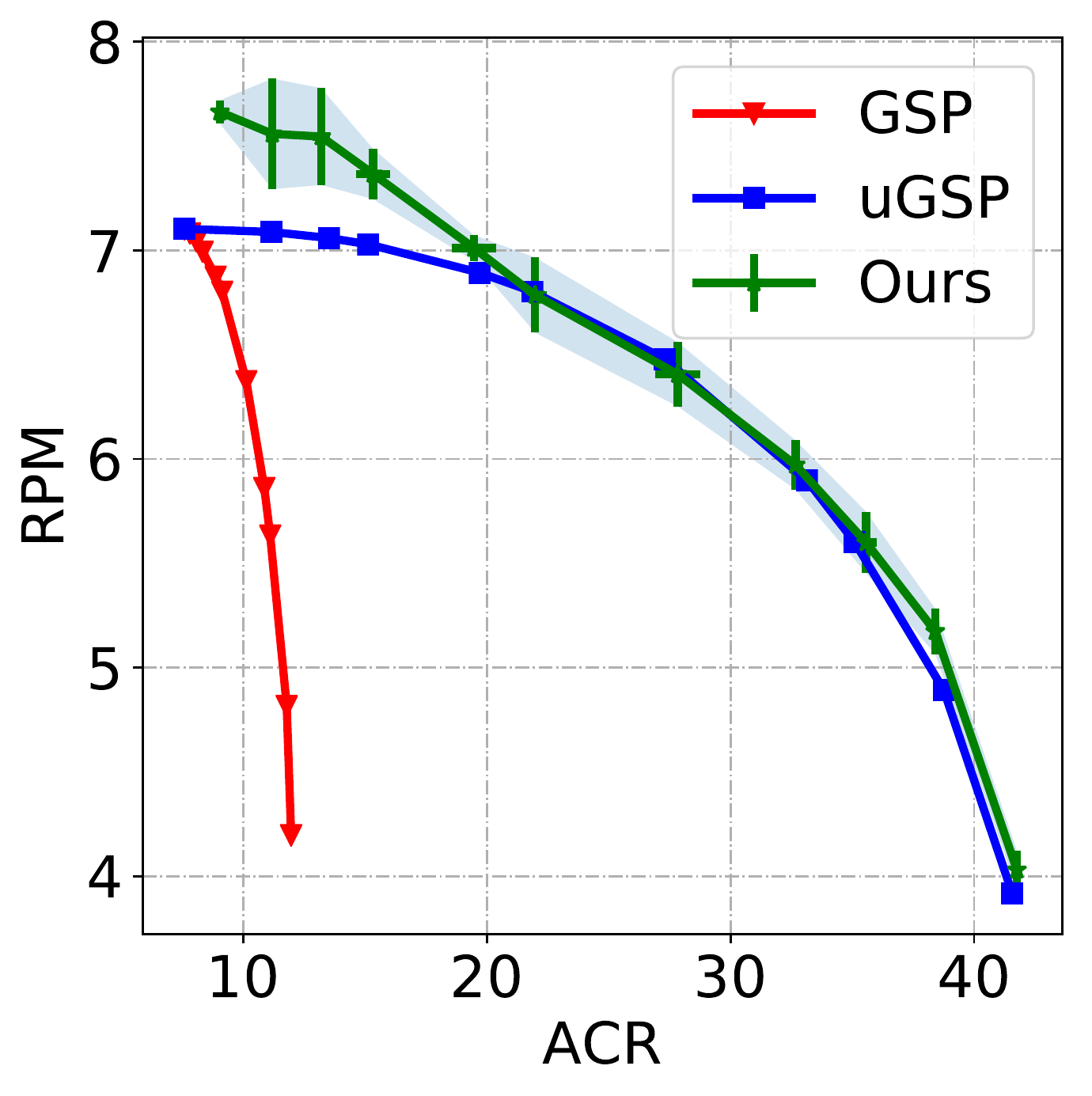}}\\
  \subfloat[CVR \& RPM]{\label{fig:offline_CVR_RPM}\includegraphics[scale=0.25]{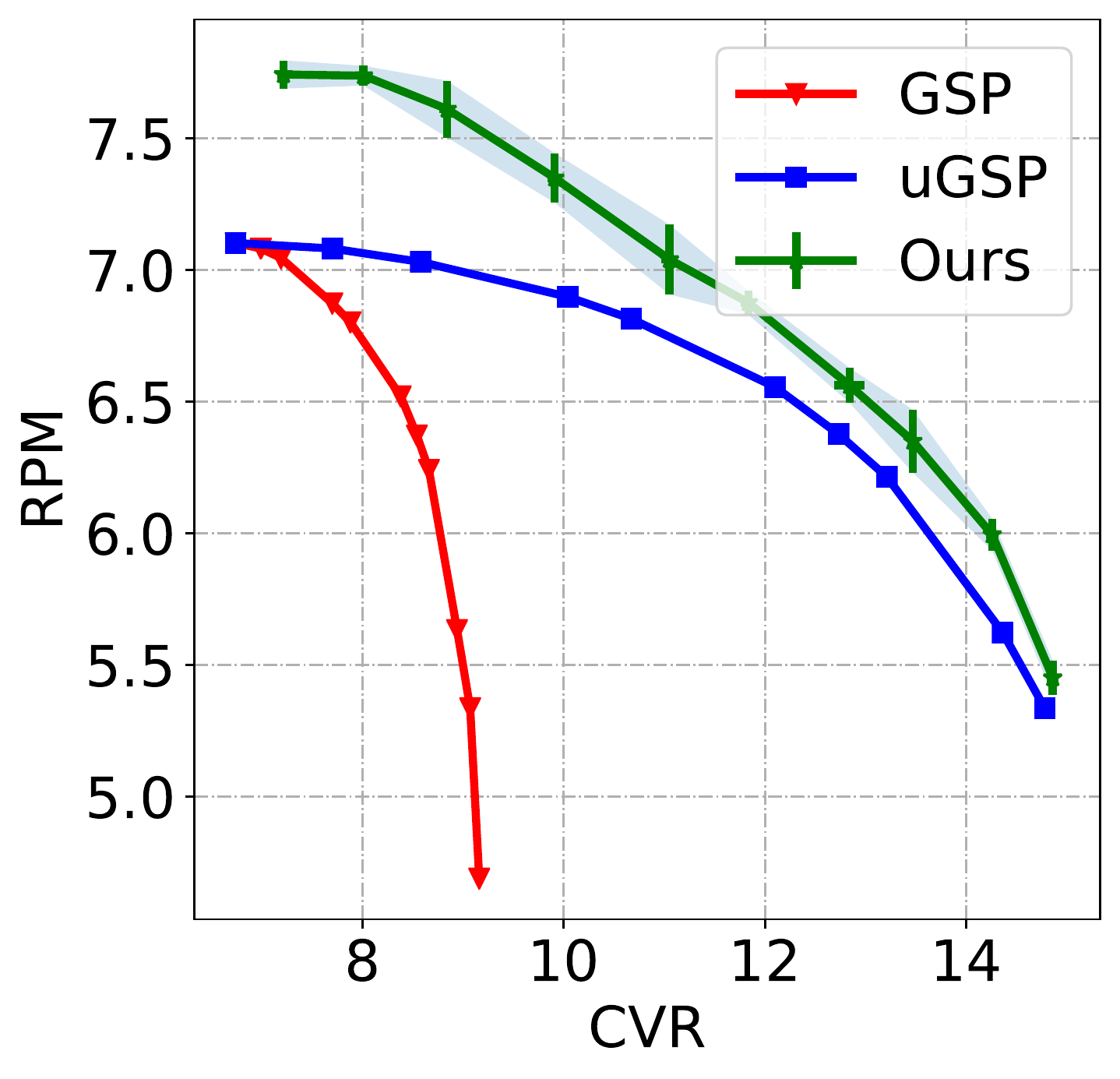}}
  \subfloat[GPM \& RPM]{\label{fig:offline_GPM_RPM}\includegraphics[scale=0.25]{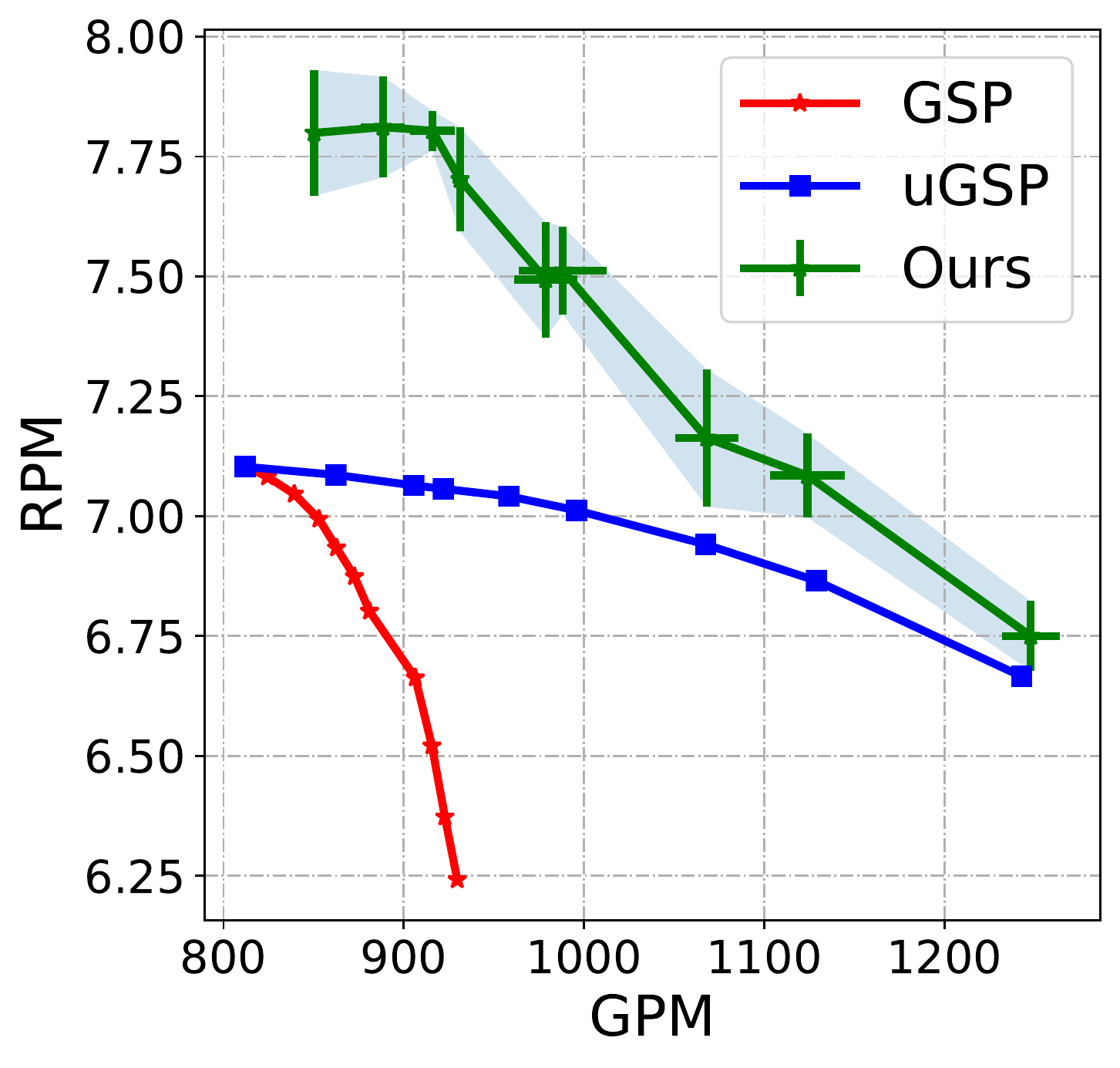}}\\
\caption{The performance of Deep GSP and other baseline mechanisms in the offline experiments.}
\label{fig:offline_total}
\end{figure}

\textbf{2) Utility-based Generalized Second Price auction (uGSP).} uGSP is a widely used mechanism in industrial ad platform, which extends the classical GSP by changing the rank score to a linear combination of more advertising objectives~\cite{bachrach2014optimising}: $r_i(b_i) = \lambda_1\times b_i\times pCTR_i + o_i$. $o_i$ represents other utilities, such as CTR and CVR: $o_i=\lambda_2\times pCTR_i + \lambda_3\times pCVR_i (\mbox{where }\lambda_t \geq 0)$. The payment of uGSP follows the principle from GSP.

\subsection{Offline Experiments}
\label{sec:exp_offline}
\subsubsection{Data sets and Offline Simulator}
\label{sec:exp_offline_simulator}
The data sets we used for experiments come from \emph{Taobao}, a leading e-commerce ad platform. We randomly select 5000k records logged data from \textit{July 4, 2020} as training data, and 870k records logged data from \textit{July 5, 2020} as test data. The logged data contains all the advertisers' bid, the estimated values ($pCTR$, $pACR$, $pCVR$, etc.), ads information (the category, the price of a product, etc.), user information (gender, age, shopping preferences, etc.), and context information (the source of traffic, etc). Due to the intractability of precise predictions, there are gaps between the estimated values (such as $pCTR$, $pACR$,  $pCVR$) and the real performance metrics (CTR, ACR, CVR). Therefore, we built an offline auction simulator with a prediction module to generate simulated feedback.

\subsubsection{Performance in Offline Simulations}
We first conduct experiments to compare the performance of Deep GSP and other baseline mechanisms without considering $\textit{ST constraint}$ (i.e., $\epsilon$ = 1.0). In order to facilitate intuitive comparisons, we set only two performance metrics with the form $\lambda \times RPM + (1-\lambda) \times X$, where $X$ is selected from $\{CTR, ACR, CVR, GPM\}$.
In Deep GSP, we directly set the objective by selecting the values of $\lambda$ uniformly from the interval $[0, 1]$. In uGSP, we set the rank score function with $\lambda \times pCTR \times bid + (1 - \lambda) \times pX$. For GSP, we tune the variable $\sigma$ in the interval $[0.5, 2.0]$. As some selected objectives may be conflicting, we plot Pareto curve of the performance metrics for different baseline mechanisms, as shown in Fig.~\ref{fig:offline_total}. Since Deep GSP utilizes a deep model with a learning algorithm, we also illustrate the error bar to demonstrate its stability.

We find that the Pareto curves of Deep GSP are mostly above the curves of other baselines, which indicates the solutions from Deep GSP outperform others. With more data fed into it, the deep rank score model can extract high-level fine-grained features and construct a more sophisticated ranking strategy to optimize the given objective, which surpasses the static mechanism (GSP, uGSP). We also notice that the performance of GSP baseline is poor when considering ACR/CVR/GPM (in Fig.~\ref{fig:offline_CR_RPM}-\ref{fig:offline_GPM_RPM}), which is reasonable as GSP does not model the effect of these indicators explicitly in its rank score function. 

\subsubsection{Monotonicity and Payment Error Rate}
\label{sec:exp_offline_mono_per_check}
\begin{figure}[!t]
\centering
\includegraphics[scale=0.25]{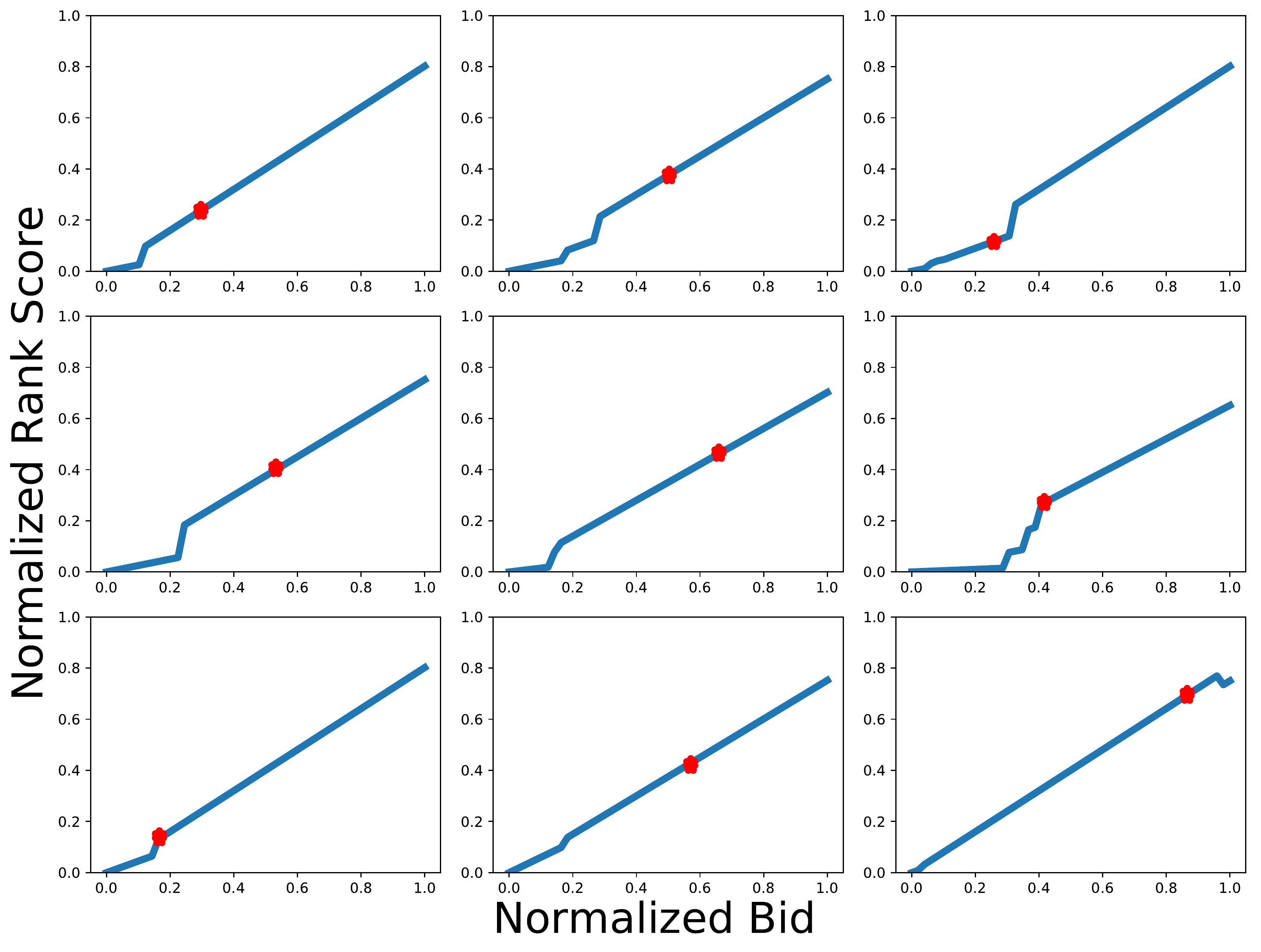}
\caption{Monotonicity Verification: Examples of the conditioned trends on bid (red markers are the real reported bids).}
\label{fig:dqn_rankscore_mono}
\end{figure}

In Fig.~\ref{fig:dqn_rankscore_mono}, we plot the trained model's conditioned trends on the bid with a few test samples. The red markers represent the real reported bids. We find that the monotonicity is guaranteed in most cases. Although some minor decreasing trends exist, we find these decreasing areas are far away from the real reported bid. As we enforce the model monotonicity by learning from data, the model may have a weak generalization on unseen state space. Table \ref{tab:offline_tabular_mono_PER} also shows the experimental results of 
deep rank score model on monotonicity metric ($\mathcal{T}_{m}$), with different performance metrics configurations. We find the $\mathcal{T}_{m}s$ are all above 0.96 on various experimental groups.
These results verify the effectiveness of the model-agnostic point-wise loss approach to guarantee monotonicity. The averaged payment errors (PER) are also given in Table~\ref{tab:offline_tabular_mono_PER}. We find all the PERs are around 1, which indicates that the approximate inverse solution $p_i$ defined in Eq.~(\ref{eq:pricing_approximated}) does not introduce much bias.

\subsubsection{Incentive Compatibility}
We now utilize the i-SIC metric~\cite{deng2020data} to evaluate \textit{incentive compatibility} (IC) of Deep GSP. The i-SIC metric is between 0 and 1, and the larger value means the better IC property. We only evaluate Deep GSP in the single-slot setting.
As we can observe from the last column in Table~\ref{tab:offline_tabular_mono_PER}, the i-SIC values are close to $1$, and the values w.r.t. different metrics configurations change negligibly. Such results demonstrate that Deep GSP can guarantee the IC property to some extent while optimizing multiple metrics, which is meaningful to benefit the long-term healthy development of the whole advertising ecology.

\begin{table}[!htbp]
    \centering
    \caption{Experimental results of deep rank score model on monotonicity, payment error ratio and IC property.\protect\footnotemark[3]}
    \small
    \begin{tabular}{|c|c|c|c|c|}
        \hline
        Exp & Metrics Configuration & $\mathcal{T}_{m}$ & PER & IC\\
        \hline
        1 & (1,0,0,0,0) & 0.991 & 1.009 & 0.9878 \\
        \hline
        2 & (0.5,0.5,0,0,0)& 0.960 & 0.994 & 0.9910 \\
        \hline
        3 & (0.5,0,0.5,0,0) & 0.978 & 0.988 & 0.9903 \\
        \hline
        4 & (0.5,0,0,0.5,0) & 0.972 & 0.995 & 0.9817 \\
        \hline
        5 & (0.5,0,0,0,0.5) & 0.982 & 0.999 & 0.9856 \\
        \hline
        6 & (0.6,0.1,0.1,0.1,0.1) & 0.975 & 0.995 & 0.9941 \\
        \hline
    \end{tabular}
    \label{tab:offline_tabular_mono_PER}
\end{table}

\begin{table}[!htbp]
    \centering
    \caption{The performance of the advertisers' utility (Adv) and platform objective (Plat) under various parameter $\epsilon$.}
    \small
    \begin{tabular}{|c|c|c|c|c|c|c|}
    \hline
        $\epsilon$ &  0.0 & 0.1 & 0.2 & 0.3 & 0.4 & $-$ST \\
         \hline
        Adv & 99.92\% & 91.71\% & 82.03\% & 69.41\% & 64.78\% & 63.95\% \\
        \hline
        Plat & 72.71\%  & 78.56\%  & 84.36\%  & 89.70\% & 96.14\% & 100\% \\
        \hline
    \end{tabular}
    \label{tab:offline_epsilon_constraint}
\end{table}

\subsubsection{Smooth Transition among Candidate Mechanisms}
Pure optimization towards the mechanism objective is not the entire goal of Deep GSP, and the smooth transition of advertisers' utility is also desired when the auction mechanism is switched. In Deep GSP, it is achieved by the $\epsilon$-constraint.
To verify the effectiveness of this constraint, we increase $\epsilon$ when the mechanism switches from RPM to CTR.
From the offline performance shown in Table~\ref{tab:offline_epsilon_constraint}, we observe that the advertiser's utility decreases in proportion to the value of $\epsilon$, illustrating that Deep GSP has the quantitative control ability towards the advertisers' performance. It is noted that the performance of both the advertisers and mechanism changes marginally when the $\epsilon$ is larger than $0.4$, due to the advertisers' strategy for guaranteeing their utilities.


\footnotetext[3]{Due to the limitation of space, we denote the weights of multiple metrics by a tuple. (e.g., for Exp2, the mechanism objective is $0.5\times RPM + 0.5\times CTR$.)}

\subsection{Online Experiments}
\label{sec:exp_online}
We present the online performance of the proposed \emph{Deep GSP auction} in \emph{Taobao} ad platform.
The deployment has the following details. (i) We use an open-source real-time log-processing framework, Flink~\cite{carbone2015apache}, to build the online stream process, which includes auction logs collection, real-time state construction, and metrics calculation. (ii) The critic and actor of the deep rank score model are trained on a Tensorflow-based distributed training framework. (iii) The exploration in policy optimization is done by deploying a separate online bucket, with random noise added on the actor output. (iv) In order to respond to the dynamic nature of the online auction environment more quickly, the deep rank score model will be updated every 15 minutes. (v) In the online engine, there are tens of thousands of traffic requests every second for the Deep GSP service, and each request contains an average of 400 ads.
After the parallelization and calculation optimization, it takes about five milliseconds for each request to be processed.

We consider five metrics, i.e., RPM, CTR, ACR, CVR, GPM, and conduct online A/B tests with several combinations of these metrics. Table~\ref{tab:online_promote} shows the online A/B test with 1\% of whole production traffic in \textit{August 1, 2020}. We use GSP as the baseline, and present the relative improvements in the table. As the online platform contains millions of user requests every day, the results can prove stable. From \textit{Exp 1 - 5}, Deep GSP makes it possible to get high performance of CTR, ACR, CVR, GPM at a low cost of RPM. From \textit{Exp 6}, we find the platform's revenue (RPM), the user experience (CTR, ACR, CVR), and overall GPM achieve a reciprocal win-win situation.

\begin{table}[!t]
    \caption{Online A/B test on different metrics configurations (August 1, 2020, 1\% production flow).}
    \centering
    \small
    \begin{tabular}{|c|c|c|c|c|c|c|}
        \hline
        Exp & Metrics & RPM & CTR & ACR & CVR & GPM\\
         \hline
        1 & RPM & \textbf{+5.2\%} & +3.1\% & -1.5\% & +0.8\% & -2.0\%\\
        \hline
        2 & RPM\&CTR & -0.3\% & \textbf{+12.8\%} & +5.6\% & +20.0\% & +7.5\%\\
        \hline
        3 & RPM\&ACR & +0.7\% & +1.5\% & \textbf{+6.6\%} & +6.8\% & +8.1\%\\
        \hline
        4 & RPM\&CVR & +0.0\% & +1.4\% & +3.6\% & \textbf{+7.5\%} & +31.0\%\\
       \hline
        5 & RPM\&GPM & +0.2\% & +3.3\% & +2.4\% & +3.6\% & +\textbf{38.7\%}\\
         \hline
        6 & All & +1.8\% & +6.2\% & +1.4\% & +5.9\% & +3.7\%\\
         \hline
    \end{tabular}
    \label{tab:online_promote}
\end{table}

\begin{figure}[!t]
\centering
\includegraphics[scale=0.35]{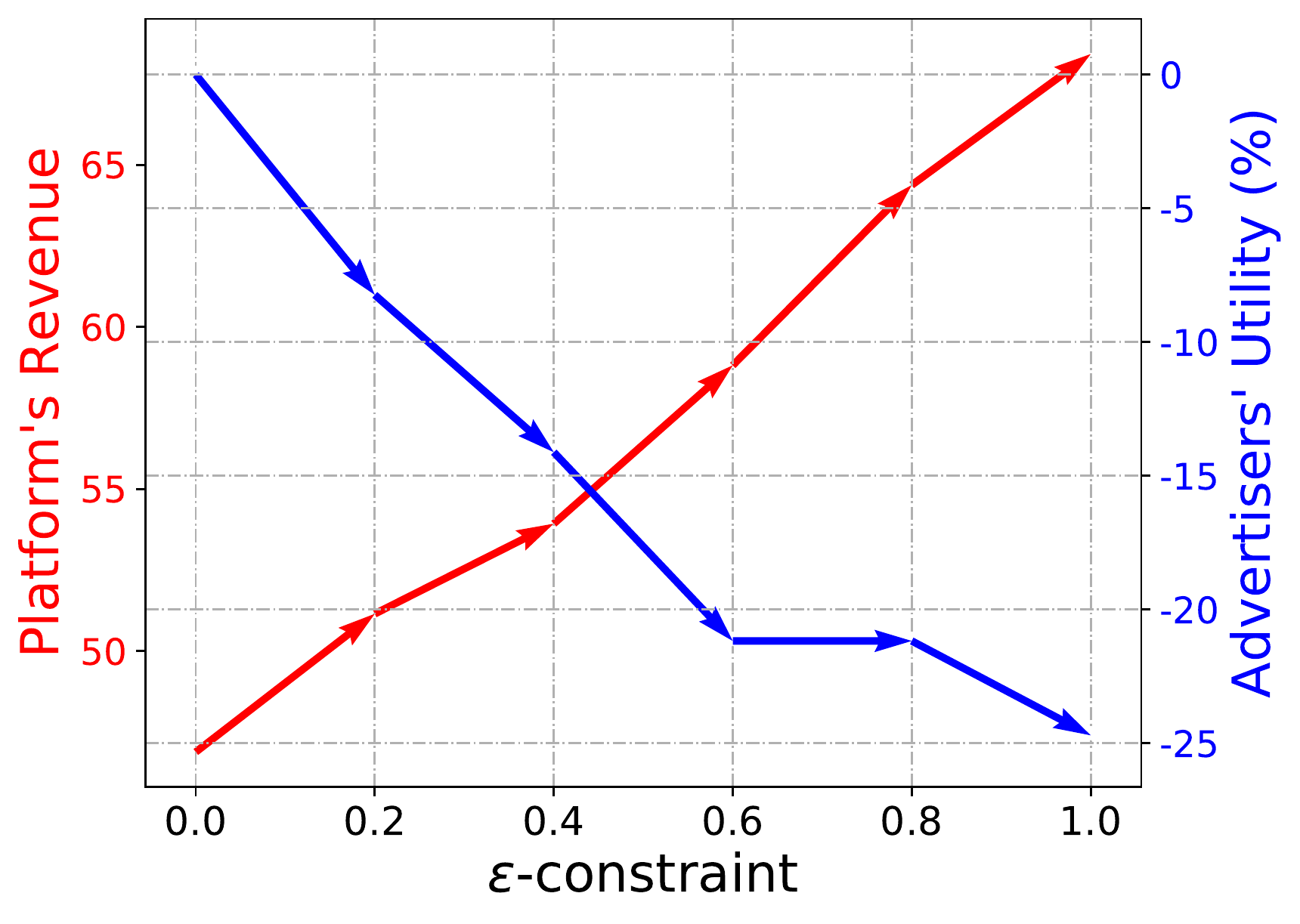}
\caption{\emph{Smooth transition} between mechanisms (from CTR to RPM) by increasing $\epsilon$ from $0.0$ to $1.0$.}
\label{fig:online_smooth_switch}
\end{figure}

Next, we verify the effectiveness of \emph{smooth transition} between auction mechanisms in the online production. We observe the influence on the advertisers' utility when the mechanism switches from CTR objective to RPM objective by adjusting $\epsilon$ from $0.0$ to $1.0$. In Fig.~\ref{fig:online_smooth_switch}, we find that the advertisers' utility (blue line) moves downward gradually along with the increase of $\epsilon$. Therefore, Deep GSP can accommodate switching between even incompatible objectives, ensuring the advertisers' performance will not immediately fluctuate too much.





\section{Related Work}
\label{sec:related_work}
Mechanism design in online advertising has been studied for a long time. The generalized second price auction (GSP)~\cite{edelman2007internet} and Vickrey-Clarke-Groves auction (VCG)~\cite{nisan2007computationally} have been widely studied and used in various advertising systems. Thompson et al.~\cite{thompson2013revenue} studied many ways to increase revenue under the GSP framework, such as reserve prices and exponential parameters. However, these works only focus on one particular optimization objective.

Several works have also discussed optimizing multiple objectives in ad auctions. \citeauthor{likhodedov2003auction}~\cite{likhodedov2003auction} proposed a framework to optimize the linear combination of revenue and social welfare in single item auctions.
\citeauthor{geyik2016joint}~\cite{geyik2016joint} discussed joint optimization of multiple performance metrics in online video advertising, including engagement, viewability, and user reach indicators. They focused on optimizing advertising campaigns and assumed the optimization objectives could be ranked in the order of importance. \citeauthor{chen2017optimizing}~\cite{chen2017optimizing} proposed a two-stage computational framework to optimize trade-offs among multiple stakeholders (platform, advertisers, and users) by incorporating various metrics. The first stage is still auction, and the second stage re-ranks ads by considering the benefits of all the stakeholders. However, their method does not explicitly consider the influence of re-ranking to the mechanism properties (such as incentive compatibility). \citeauthor{bachrach2014optimising}~\cite{bachrach2014optimising} proposed truthful auction mechanisms to optimize trade-offs between multiple stakeholders.
They designed rank score function as a linear combination of revenue, welfare, and clicks, while the payment was computed as prescribed by \citeauthor{myerson1981optimal}~\cite{myerson1981optimal}.
Their method requires accurate model prediction of each metric when applied to optimize multiple performance metrics in real industrial applications.

Another concurrent line of work studied by~\cite{conitzer2004self,sandholm2003automated} is automated mechanism design.
They framed the problem of optimizing revenue and welfare as an instance of a constraint satisfaction problem (CSP) and solved CSP with the linear program (LP) algorithms.
 \citeauthor{duetting2019optimal}~\cite{duetting2019optimal} used deep learning in the context of mechanism design under incentive compatibility constraints or low ex-post regret. They formulated the optimal auction design as a constrained optimization problem, without consideration of actual feedback.
 \citeauthor{shen2017reinforcement}~\cite{shen2017reinforcement} and \citeauthor{tang2017reinforcement}~\cite{tang2017reinforcement} modeled the impression allocation problem as a Markov decision process and solved it by reinforcement learning. However, most of the above mechanisms are built on optimizing single optimization objective, such as revenue and social welfare.

\section{Conclusion}
\label{sec:conclusions}
In this paper, we focus on the problem of optimizing multiple performance metrics in online e-commerce and propose an end-to-end learning based ad auction mechanism. We leverage the deep learning technique to design a new rank score function and integrate it into the GSP auction framework, i.e., Deep GSP auction. 
We also mathematically characterize the optimization of multiple performance metrics under some desirable properties in Deep GSP auction, such as \emph{game equilibrium} and \emph{smooth transition}, and give detailed algorithms with other relative technical details such as point-wise monotonicity loss, approximate inverse payment, and deep policy optimization. Extensive experiments have been conducted on a real-world e-commerce ad platform. Both offline and online experimental results validate the effectiveness of the proposed auction mechanism.

\begin{acks}
This work was supported in part by Science and Technology Innovation 2030 – ``New Generation Artificial Intelligence'' Major Project No. 2018AAA0100900, in part by Alibaba Group through Alibaba Innovation Research Program, in part by China NSF grant No. 61902248, 62025204, in part by Shanghai Science and Technology fund 20PJ1407900. The authors would like to thank Han Li, Xiaoqiang Zhu, Yu Rong, Hongtao Lv, Junqi Jin, Rihan Chen, Rui Du, Guan Wang and anonymous reviewers for their valuable help and suggestions.
\end{acks}

\bibliographystyle{ACM-Reference-Format}
\bibliography{sample-base}

\end{document}